\documentclass[aps,prb, reprint, citeautoscript, showpacs,floatfix,superscriptaddress]{revtex4-2}

\usepackage{bm}
\usepackage{amssymb}
\usepackage{graphicx}
\usepackage{amsmath}

\begin{document}
\title{Excitation Energy Transfer in Nanohybrid System of Organic Molecule and Inorganic Transition Metal Dichalcogenides Nanoflake}

\author{Yan Meng}
\affiliation{Department of Physics, University of Science and Technology Beijing, Beijing, 100083, China.} 

\author{Kainan Chang}
\email{knchang@ciomp.ac.cn}
\affiliation{Department of Physics, University of Science and Technology Beijing, Beijing, 100083, China.}

\author{Luxia Wang}
\email{luxiawang@sas.ustb.edu.cn}
\affiliation{Department of Physics, University of Science and Technology Beijing, Beijing, 100083, China.}

\begin{abstract}
Excitation energy transfer (EET) in an organic/inorganic nanohybrid system, 
composed of a single \textit{para}-sexiphenyl (6P) molecule physisorbed on a  finite-sized MoS$_2$ nanoflake,  is investigated theoretically. 
The electronic structure of the MoS$_2$ nanoflake is described by using an 11-band tight-binding model, in which edge states are passivated with H atoms to restore a well-defined bandgap.
Within a configuration-interaction scheme, excitonic states are constructed and, for computational efficiency, approximated by uncorrelated electron-hole pairs in the relevant high-energy window.
The EET rates are evaluated via Fermi's golden rule, incorporating Coulomb coupling, thermal broadening, and spectral overlap between the molecular excitation and the MoS$_2$ nanoflake's electron-hole pairs.
Our results reveal that  energy transfer from the molecule to the nanoflake is the dominant process, and its efficiency depends strongly on the size of the MoS$_2$ nanoflake, as well as the molecule's vertical distance and lateral position  relative to the nanoflake.

\end{abstract}
\maketitle

%
%

\section{Introduction}

Two-dimensional (2D) transition-metal dichalcogenides (TMDCs) are a class of layered semiconducting materials formed by a combination of transition-metal atoms (Mo, W, etc.) and chalcogen atoms (S, Se, Te, etc.).
They exhibit high carrier mobility,  direct-to-indirect bandgap crossover with increasing layer numbers, and strong exciton binding energy \cite{Mak-2010-p136805, Splendiani-2010-p1271,Raja-2017-p15251,Yuan-2017-p3371},  which collectively yield unique electronic and optical properties.
Therefore, TMDCs have attracted considerable 
interest for their potential applications
in next-generation nanoscale electronics and photonics \cite{Obaidulla2024, Ahmed2024, Lin-2014-p5569,Novoselov-2005-p10451,Mak-2016-p216,Zong-2008-p7176}.
Despite these advantages, most TMDCs suffer from narrow absorption bands, poor light absorptions, and high production costs \cite{Li2020}. 
To expand their functionality, TMDCs are often integrated with other materials into heterostructures, such as organic molecules \cite{Huang2018,Huang2018a,Greulich2020,Mutz-2020-p2837,Mandic2026}, organic semiconductors \cite{Markeev2022,Gu-2018-p100}, quantum dots \cite{Tang2020,Raja2016}, polymers \cite{Shastry2016,Zhong2018}, and perovskites \cite{Fang-2018-p1655,Gillespie2024}.

Among these, organic/inorganic nanohybrid systems have shown promise in devices like photodetectors, solar cells, and light-emitting diodes \cite{Huang2018a,Mandic2026}.
Organic components offer additional benefits -- including low cost, broad absorption bandwidth, durability, flexibility, light weight, and easy processing \cite{Li2020,Amsterdam-2019-p4183} -- 
making them ideal for improving carrier mobility and photoluminescence yield in TMDCs. 
Currently, the understanding of the physical mechanism 
of carrier mobility is based mainly on the alignment of energy level at the interface between organic molecules and TMDCs, resulting in  surface charge transfer \cite{Huang2018a,Amsterdam-2019-p4183,Mouri2013}.
Theoretically, to realize the surface charge transfer, it requires the overlap of wave functions between the two materials, as well as the charge exchange contributions. 
However, even in the absence of energy level alignment or wave function overlap,  long-range Coulomb interactions
can induce the energy exchange via a process known as the excitation energy transfer (EET) or  F{\"o}rster resonance energy transfer (FRET) \cite{Mandic2026}, which cannot be ignored in the optical process of such heterostructures.

\begin{figure}[ht] 
	\centering   
	\includegraphics[scale=0.29]{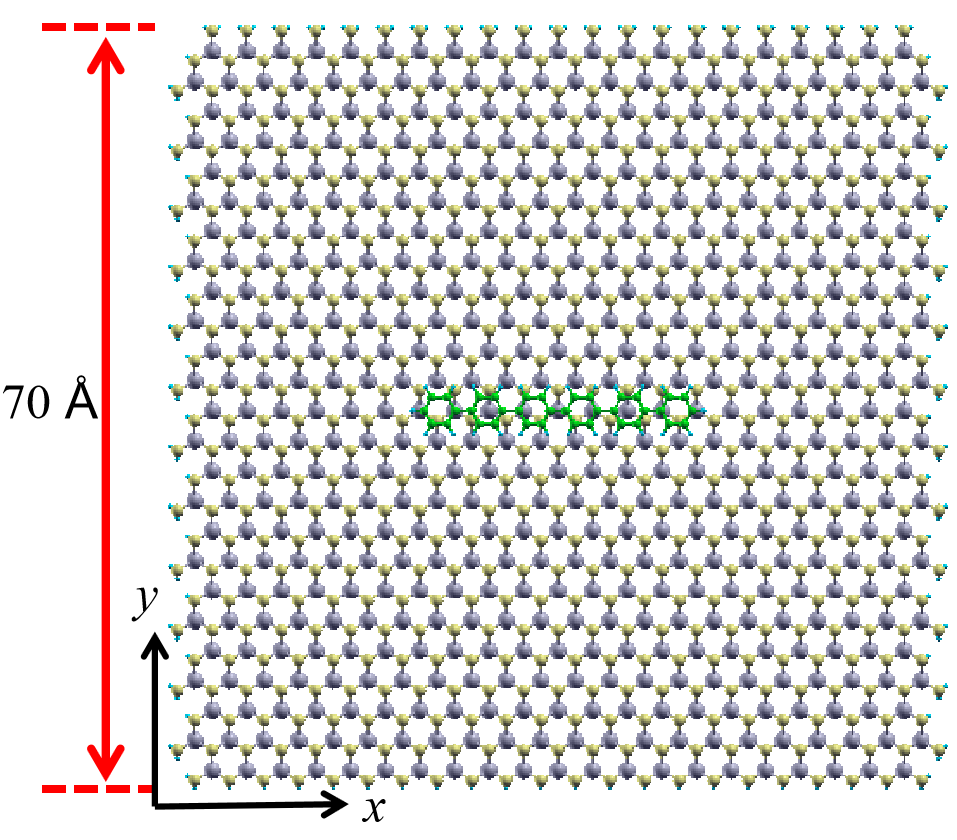}  
	\includegraphics[scale=0.29]{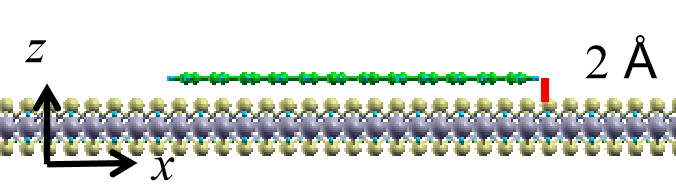}  
	\\
	\includegraphics[scale=0.29]{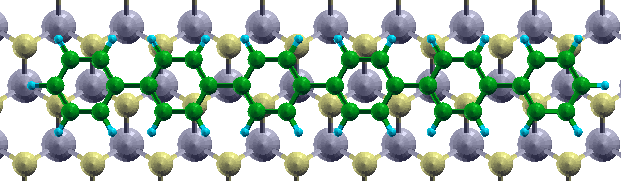}  
	\caption{Schematic illustration of nanohybrid system formed by the H-passivated MoS$_2$ nanoflake and the 6P molecule. The upper part shows the top and the side views, while the lower part displays an enlarged top view (yellow sphere: S, gray sphere: Mo, cyan sphere: H, green sphere: C).
$x$, $y$, $z$ denote Cartesian coordinates. } 
	\label{structure} 
\end{figure}

Experimentally, TMDC/organic heterostructures are commonly
fabricated through various physical and chemical strategies \cite{Huang2018,Huang2018a}, where organic molecules form a layer covering the surface of TMDC.
Focusing on the EET mechanism, theoretically, it is reasonable to study a system comprising a single molecule and a monolayer TMDC nanoflake,
ensuring that the TMDC nanoflake is significantly larger in size than the molecule.
In this work, we select 
a well-studied material MoS$_2$ as a representative TMDC,  constructing  a finite square nanoflake. Its edges are terminated by unsaturated S atoms, which are more stable than the Mo edges \cite{Javaid-2017-p2045, Jeppe-2007-p53,  Helveg-2000-p951}.
Furthermore,  to widen the bandgap, the unsaturated S atoms are passivated with H atoms \cite{Loh-2015-p1565,Javaid-2017-p2045,Arul2016}, thereby removing dangling bonds.
We calculate the electronic structure of the H-passivated MoS$_2$ nanoflake by using an 11-band tight-binding  model \cite{Cappelluti}, with coupling parameters involving H and other atoms assigned manually.
The organic component is chosen as a \textit{para}-sexiphenyl (6P) molecule, which can be synthesized and purified in a well-defined manner \cite{Zojer-2000-p16538,Sparenberg-2014-p26084} and has been well integrated with 2D materials \cite{Hlawacek2011,Hlawacek-2011-p333,Sun-2019-p1803831}.
Owing to its planar geometry, the 6P molecule naturally lies flat on a plane above the MoS$_2$ surface.
Figure\,\ref{structure} illustrates the nanohybrid system formed by the MoS$_2$ nanoflake and the 6P molecule.
As an example,  the  MoS$_2$ nanoflake is modeled as a square with a side length of
70 {\AA} with 1703 Mo and S atoms, which is larger than the length of 6P molecule around 27 {\AA}.
Additional nanoflake sizes -- 60 {\AA} and 74 {\AA}, containing 1177 and 1853 atoms, respectively -- are also examined.
Here, the 6P molecule is capable of moving within a plane situated above the surface of MoS$_2$ while maintaining a minimum vertical separation of 2 {\AA}, which prevents the overlap of wave functions. Besides, the excitation energy of 6P molecule is about 4 eV \cite{Wang2020}, which is much larger than the bandgap of MoS$_2$, thus the energy level alignment disappears.
Under these conditions, Coulomb interactions between the MoS$_2$ nanoflake and 6P molecule lie in the millielectronvolt range or less (see below), which motivates a Fermi's golden rule description of EET rate expressions \cite{May}.
Besides, the EET process is expected to depend strongly on the relative position and orientation of the molecule relative to the nanoflake, which will be explored here.
Compared with first-principles approaches relying on periodic boundary conditions or requiring phenomenological parameters to describe the coupling, our approach combines an efficient tight-binding model for the electronic structure of the large-scale MoS$_2$ nanoflake (thousands of atoms) with an explicit calculation of Coulomb integrals based on atomic transition charges.
This framework directly yields the distance-dependent coupling strength without introducing fitted parameters and enables a systematic investigation of how the lateral position influences the EET efficiency.

This paper is organized as follows:
In Sec.\,\ref{sec2}, we present the  theoretical framework for
 EET process in the TMDC/6P heterostructure, including a description of electronic structure and exciton states of MoS$_2$ nanoflake, followed by the derivation of the EET rate expression.
In Sec.\,\ref{sec3}, we show the results and discussions, beginning with the effect of H passivation and electron-hole pair approximation. This foundation enables the calculation and analysis of  EET rate for various configurations.
Finally, in Sec.\,\ref{sec4}, we briefly summarize this work.

%
%

\section{Theory}
\label{sec2}

\subsection{Electronic structure of MoS$_2$ nanoflake}

An 11-band tight-binding model, which has been verified as an effective approach for describing the electronic structure of 2D TMDCs \cite{Cappelluti}, is employed to characterize the MoS$_2$ nanoflake in this work. 
 Within each unit cell, the basis set consists of
 five $d$ orbitals from the Mo atom and six $p$ orbitals from the S atom (the three $p^t$ and three $p^b$ orbitals for the top layer and bottom layer S atom, respectively).
This configuration constructs an 11-dimensional Hilbert space with this orbital basis:
\begin{align}
\phi=\left( p_x^t, p_y^t, p_z^t,d_{3z^2 - r^2},d_{xz}, d_{yz},d_{x^2-y^2},
d_{xy},p_x^b, p_y^b, p_z^b\right)\,.
\label{11-band-basis}
\end{align}
Thus, the  real-space tight-binding Hamiltonian for the finite MoS$_2$ nanoflake  is then expressed as  \cite{Peng2024}
\begin{align}
H = \sum_{i,\mu \nu} \epsilon_{\mu\nu}c^\dagger_{i,\mu}c_{i,\nu}
+\sum_{i\neq j,\mu\nu}
\left[ 
t_{ij,\mu\nu} c^\dagger_{i,\mu}c_{j,\nu} + \text{H.c.}
\right]\,,
\end{align}
with the creation (annihilation) operator $c^\dagger_{i,\mu}$ ($c_{i,\nu}$)  for the electron in orbital $\mu$ ($\nu$) of Eq.\,\eqref{11-band-basis}  in unit cell $i$,  the on-site energies $\epsilon_{\mu\nu}$, and  the hopping integrals $t_{ij,\mu\nu}$ between orbitals $\mu$ and $\nu$ at sites $i$ and $j$.
Here, $\epsilon_{\mu\nu}$ are related to the crystal field ($\Delta_0$, $\Delta_1$, $\Delta_2$ for $d$ orbitals of Mo atom, $\Delta_p$, $\Delta_z$ for $p$ orbitals of S atom), and $t_{ij,\mu\nu}$ are determined by the Slater-Koster parameters ($V_{pd\sigma}$, $V_{pd\pi}$ for Mo--S bonds, $V_{dd\sigma}$, $V_{dd\pi}$, $V_{dd\delta}$ for Mo--Mo bonds, $V_{pp\sigma}$, $V_{pp\pi}$ for S--S bonds), as established in previous literature \cite{Cappelluti,Roldan-2014-p34003,Peng2024}.

For a finite system containing $N$ unit cells, the Hamiltonian matrix dimension is $11N\times11N$.
Diagonalization yields the single-particle states,
 expanded in the local atomic orbital basis as
\begin{align}
\varphi_a(\mathbf{r}) = \sum_{i,\mu} C_{i\mu,a} \phi_{i\mu}(\mathbf{r}-\mathbf{R}_i)\,,
\label{wavefunction}
\end{align} 
where $C_{i\mu,a}$ are expansion coefficients and $\phi_{i\mu}$ denotes the orbital  $\mu$  at site $i$.
The resulting energy spectrum contains $11N$ discrete levels $E_a$, which, 
in the ground state, can separate into 
the hole states with energies $E_{\bar{a}}$ corresponding to occupied valence bands, and the electron states with energies $E_{a}$ representing unoccupied conduction bands.
Note that spin-orbit coupling is neglected in the present formulation, leaving the Hamiltonian spin-degenerate.


%
%

\subsection{Exciton states of MoS$_2$ nanoflake}

Within  a configuration interaction scheme,
the exciton states of the nanoflake are constructed from the single-particle electron and hole states,  incorporating their mutual Coulomb attraction. 
In second quantization, the excitonic Hamiltonian reads \cite{Ziemannn-2015-p4054,Plehn-2015-p7467}
\begin{align}
\label{WMX}
H_{\text{X}}= \sum_a E_a e_a^\dagger e_a - \sum_{\bar{a}} E_{\bar{a}}  h_{\bar{a}} ^\dagger h_{\bar{a}} 
+\sum_{a,b} \sum_{{\bar{a}},{\bar{b}}} V_{a{\bar{a}} ,{\bar{b}}b }^{(e-h)} e^\dagger_{a} h^\dagger_{\bar{a}} h_{\bar{b}} e_b \,,
\end{align}
with operators $e^\dagger_a$ and $e_a$ for electrons, and $h^\dagger_{\bar{a}}$ and $h_{\bar{a}}$ for holes.
The electron-hole coupling term splits into two contributions, as
\begin{align}
\label{coupling}
V^{(e-h)}_{a{\bar{a}} ,{\bar{b}}b} = - W^{(e-h)}_{a{\bar{a}} ,{\bar{b}}b} + J^{(e-h)}_{a{\bar{a}} ,{\bar{b}}b} \,,
\end{align}
constituted by the Coulomb matrix element
\begin{align}
\label{w}
W^{(e-h)}_{a {\bar a}, {\bar b} b} 
= \sum_{u, v} \Big( \delta_{u, v} V_0/e^2 
+\frac{1 - \delta_{u, v}}{\epsilon|{\bf R}_u - {\bf R}_v |} \Big)
q_u(a b) q^*_v({\bar a} {\bar b})\,,
\end{align}
and the exchange matrix element 
\begin{align}
\label{j}
J^{(e-h)}_{a {\bar a}, {\bar b} b} 
= \sum_{u, v} \Big( \delta_{u, v} V_0/e^2 
+\frac{1 - \delta_{u, v}}{\epsilon|{\bf R}_u - {\bf R}_v |} \Big)
q_u(a {\bar a}) q^*_v(b {\bar b})\,.
\end{align}
The dielectric constant of MoS$_2$  is taken as $\epsilon=6.8$ from Ref.\,\citenum{Butoi-1998-p9635}, with a distance-dependent screening function following Ref.\,\citenum{Korkusinski-2010-p245304}.
The on-site term of the Coulomb coupling is denoted as $V_0/e^2$;
and it can be evaluated by introducing spherical coordinates and  expanding in spherical harmonics, in which Wigner's 3j symbols are used to compute integrals over three spherical harmonics.
Besides, $q_u(ab)$, $q_u(\bar{a}\bar{b})$, and $q_u(a\bar{a})$ (or $q_u(b\bar{b})$) are transition charge densities between electron-electron, hole-hole, and electron-hole, respectively, which are regarded as localized on position ${\bf R}_u$ of atom $u$ in the atomic partial charge approximation \cite{Ziemannn-2015-p4054}.


Due to two or more electron-hole pairs being of less interest in this work, electron-electron and hole-hole Coulomb couplings are neglected.
Therefore, the single-exciton states formed by electron-hole pair configurations are written as
\begin{align}
\label{x-states}
|\alpha\rangle = \sum\limits_{a, {\bar a}}  C_{\alpha}(a {\bar a})|\psi_{a {\bar a}}\rangle\,,
\end{align}
where electron-hole pair states are denoted as
$|\psi_{a {\bar a}}\rangle = e^+_a h^+_{\bar a} |\psi_0\rangle$
with the ground states $|\psi_0\rangle$.
The excitonic energies $\mathcal{E}_\alpha$ and the corresponding expansion coefficients $C_{\alpha}(a {\bar a})$ are obtained by solving the following equation:
\begin{align}
\big(\mathcal{E}_\alpha -E_a + E_{\bar a} \big)C_{\alpha}(a {\bar a})
=- \sum\limits_{b, {\bar b}} \big( W^{(e-h)}_{a {\bar a}, {\bar b} b}
- J^{(e-h)}_{a {\bar a}, {\bar b} b} \big) C_{\alpha}(b {\bar b})\,.
\label{exciton}
\end{align}

\subsection{Expression for Rates of EET}

For the present case, it is sufficient to consider rates of EET between the 6P molecule and the MoS$_2$ nanoflake.
Within a generic molecular donor-acceptor framework, the EET rate can be represented via Fermi's golden rule as $k_{\rm EET} = 2\pi/\hbar \times |V_{DA}|^2 \mathcal{D_{\rm EET}}$. 
Here, $V_{DA}$ describes the transition Coulomb interaction between the donor (D) and the acceptor (A),
and 
$\mathcal{D_{\rm EET}} = \int {\rm d}E \mathcal{D^{\rm (D)}_{\rm e \to g}}(E)\mathcal{D^{\rm (A)}_{\rm g \to e}}(E) $ is the combined density of states (DOS) referring to the de-excitation of the donor $\mathcal{D^{\rm (D)}_{\rm e \to g}}(E)$ and the excitation of the acceptor $\mathcal{D^{\rm (A)}_{\rm g \to e}}(E)$ \cite{Ziemann-2014-p1203}.

Hence, for the EET process from the MoS$_2$ excitonic levels to the single 6P molecule can be characterized by
\begin{align}
k_{\rm sem \to mol}=
\frac{2\pi}{\hbar}\sum_\alpha  f_{\alpha}|V_{e0,\alpha g}|^2 
\mathcal{D}^{\rm (A)}_{g \to e}(\mathcal{E}_\alpha)\,.
\label{eq-rate2}
\end{align}
In the molecular part,
$\mathcal{D}^{\rm (A)} _{g\to e}$ is determined by the overlap between the semiconductor exciton energies $\mathcal{E}_\alpha$ and the molecular excitation spectrum, and can be future approximated by a standard high-temperature version
\begin{align}
\label{D1}
\mathcal{D}^{\rm(A)} _{g\to e}(\mathcal{E}_\alpha) 
= \frac{1}{\sqrt{2\pi k_B T S_{eg}} } {\rm exp} \left\{ -\frac{(\mathcal{E}_\alpha-E_{eg}-S_{eg}/2)^2}{2 k_B T S_{eg}}\right\}\,,
\end{align}
with the so-called 0-0 transition energies $E_{eg}$ and the Stokes shift $S_{eg}$.
While in MoS$_2$ part, $\mathcal{D^{\rm (D)}_{\rm e \to g}}(E)$ is simply given by $\delta(E -\mathcal{E}_\alpha)$,  and $f_{\alpha}$ denotes the thermal distribution of the relaxed semiconductor excitations.
Their EET coupling $V_{e0,\alpha g}$ is defined by the Coulomb interaction between the molecular transition charge density $\rho _{eg}$ and the transition charge density $\rho_{\alpha 0}$ of the semiconductor. 
We introduce atomic-centered partial charges as $\rho _{eg} \approx \sum _u q_u(eg)\delta (\mathbf{x}-\mathbf{R}_u)$ and $\rho_{\alpha 0} \approx \sum_v q_v(\alpha 0)\delta (\mathbf{y}-\mathbf{R}_v)$.
Hence, the transition coupling is denoted as
\begin{align}
\label{x-coupling}
V_{e0,\alpha g}= \sum_{u,v} \frac{q_u (eg) q^*_v(\alpha 0)}{|\mathbf{R}_u - \mathbf{R}_v|} \,.
\end{align}
Here, the atomic-centered transition charges of the 6P molecule is $q_u (eg)$. 
The exciton transition charges of the semiconductor $q^*_v(\alpha 0)$ are dependent on the electron-hole transition charges $q_v(a \bar{a}) = e\sum_o C^*_a(vo)C_{\bar{a}}(vo)$ as
\begin{align}
\label{transition-charges}
q_v(\alpha 0) &= \sum_{a,s, \bar{a},\bar{s}} C^*_\alpha(as, \bar{a}\bar{s}) q_v(a \bar{a})
\notag \\
&=e \sum_{a,s, \bar{a},\bar{s}} C^*_\alpha(as, \bar{a}\bar{s})
\sum_o C^*_a(vo) C_{\bar{a}}(vo)\,.
\end{align}

For the reverse EET from the single 6P molecule to the MoS$_2$ nanoflake,
the rate is given by
\begin{align}
k_{\rm mol \to sem} = \frac{2 \pi}{\hbar} \sum _\alpha  |V_{g \alpha, 0 e }|^2 \mathcal{D} _{e\to g}(\mathcal{E}_{\alpha}) \,,
\label{eq-rate}
\end{align}
with 
\begin{align}
\label{D2}
\mathcal{D}^{\rm (D)} _{e\to g} (\mathcal{E}_\alpha)
=\frac{1}{\sqrt{2\pi k_B T S_{eg}} } {\rm exp} \left\{ -\frac{(\mathcal{E}_\alpha-E_{eg}+S_{eg}/2)^2}{2 k_B T S_{eg}}\right\}\,.
\end{align}
Note, $V_{g \alpha, 0 e }$ and $V_{e0,\alpha g}$ are identical.

%
%

\section{Results and Discussions}
\label{sec3}

In the following, we set the room temperature $T=300$ K, the 6P molecule excitation energy $E_{eg}=4.014$ eV \cite{Plehn-2018-p27925}, with  Stokes shift $S_{eg}=0.022$ eV \cite{Ziemann-2014-p1203}.

\subsection{Edge state passivation}

\begin{figure*}
       \centering  
       \begin{minipage}{0.4\linewidth}
         \includegraphics[scale=0.18]{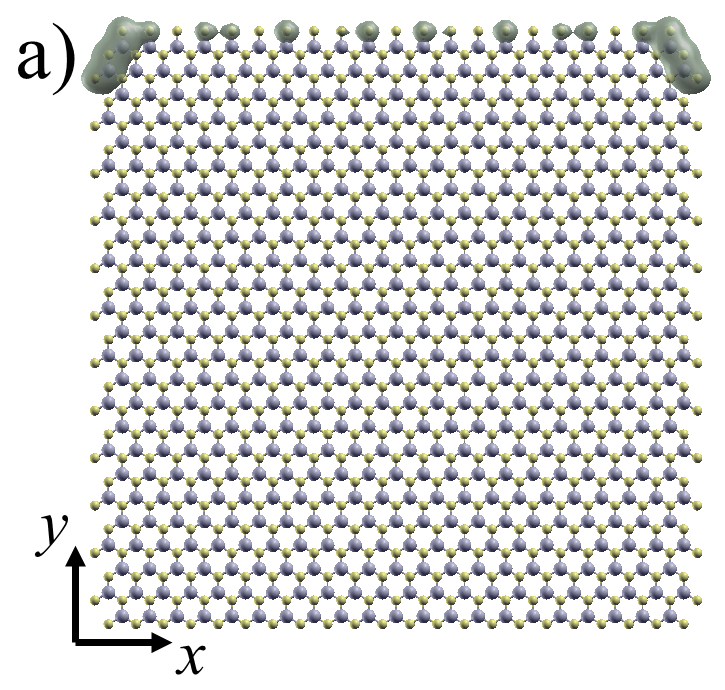}
         \includegraphics[scale=0.18]{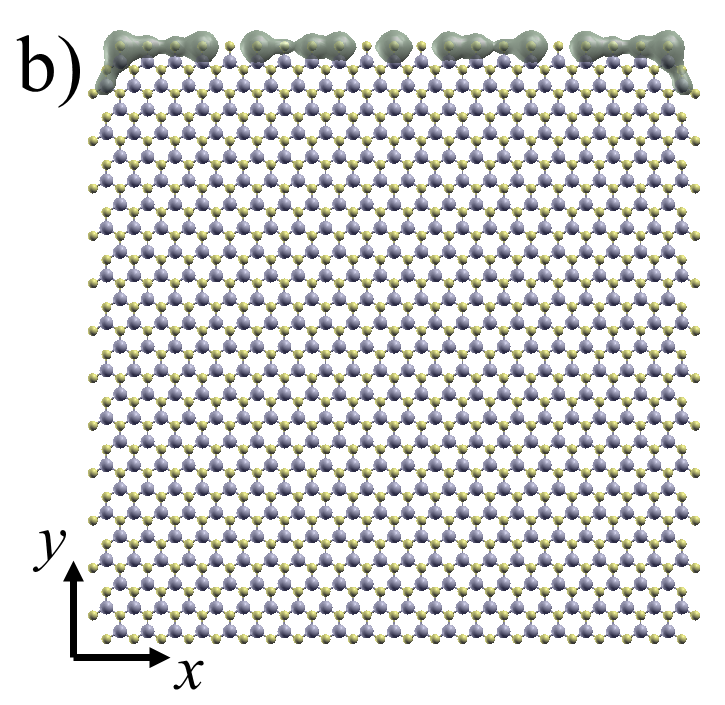}

         \includegraphics[scale=0.18]{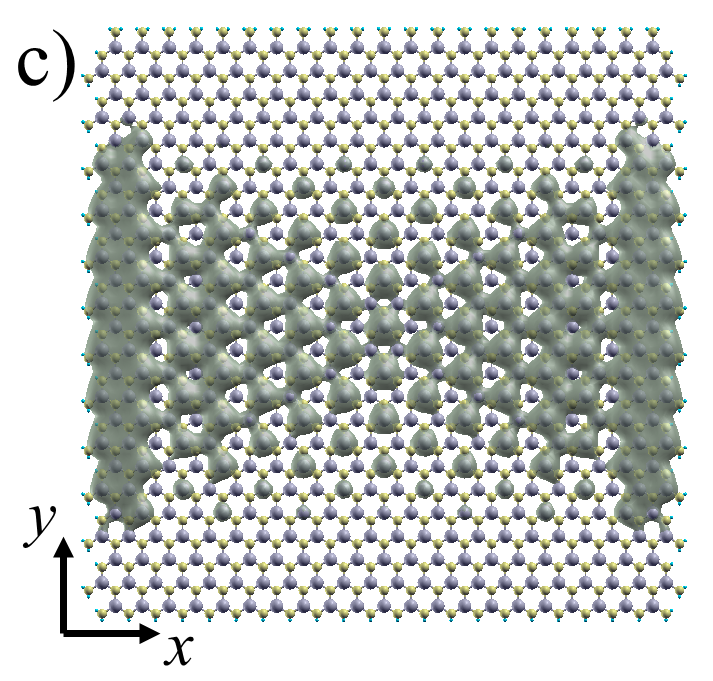}
         \includegraphics[scale=0.18]{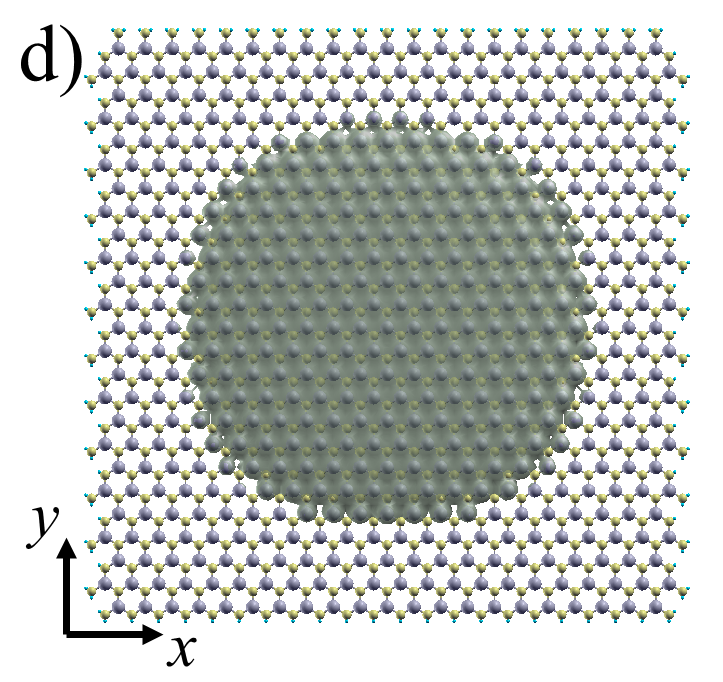}
       \end{minipage}
       \begin{minipage}{0.59\linewidth}
            \includegraphics[scale=1]{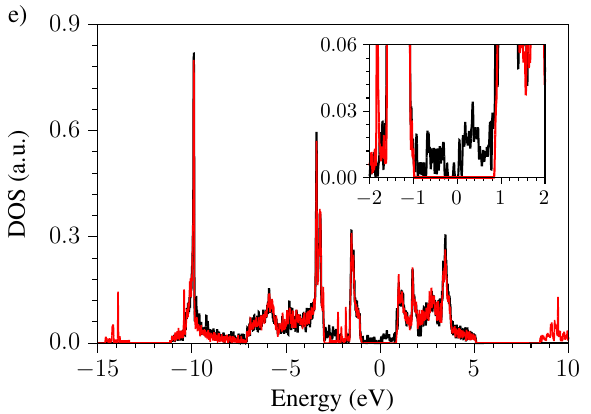} 
       \end{minipage}
       \caption{Density of single-electron states $|\varphi_a ({\bf r})|^2$ referring to unpassivated (a) the conduction-band like state and (b) valence-band like state, as well as (c,d) the H-passivated cases.
       (e) Total DOS for unpassivated (black solid line) and passivated (red dashed line) cases. }
       \label{LH}
\end{figure*}

Although a MoS$_2$ monolayer exhibits a bandgap of approximately 1.84 eV, the finite nanostructure will introduce unphysical surface states inside the bandgap, i.e., 
the so-called in-gap states or edge-states \cite{Kilin-2007-p342,Pav2015}.
In order to remove these states directly, the H atoms are placed next to the edge S atoms and the interatomic H-S distance is about 1 {\AA}.
The effect of H-passivation on the electronic states can be visualized as follows (taking 70 {\AA} MoS$_2$ nanoflake as an example):
(1) By calculating $|\varphi_a ({\bf r})|^2$, we obtain the charge density related to single-electron states in Fig.\,\ref{LH}. 
Without passivation, the conduction band (CB)-like states and valence band (VB)-like states are shown in Figs.\,\ref{LH}\,(a) and (b), respectively.
These states are  found to strongly localize at the edges.
In the presence of the H-passivation, these states
become distributed predominantly in the central region of the nanoflake, as shown in Figs.\,\ref{LH}\,(c) and (d).
Clearly, contributions from edge atoms have been removed, that is, the passivation of the dangling bonds effectively excludes the boundary effects.
This occurs because H atoms saturate the dangling bonds of edge S atoms, eliminating the localized electronic states within the bandgap that are artifacts of the finite boundary. 
Consequently, the electronic structure of the passivated nanoflake more closely resembles that of an infinite 2D sheet, with a well-defined band gap.
(2)
The normalized DOS of the MoS$_2$ nanoflake also quantifies this effect in Figs.\,\ref{LH}\,(e).
The unpassivated MoS$_2$ nanoflake has no bandgap around zero energy, while after the passivation, a clear band gap of about 1.9 eV can be observed (see the inset).
Meanwhile, the added H atoms introduce the artificial states below $-11$ eV and above 5 eV, which lie far from the relevant energy window and do not significantly affect the electronic transitions of materials.
In this way, we achieve the application of the 11-band tight-binding model to the calculation of electronic states in finite TMDC nanoflake, and obtain the correct bandgap by passivating the boundary S atoms through H atoms.
Besides, as the system size increases, this gap decreases until that of a periodic structure.

\begin{figure}[ht]
      \centering  
\includegraphics[scale=0.23]{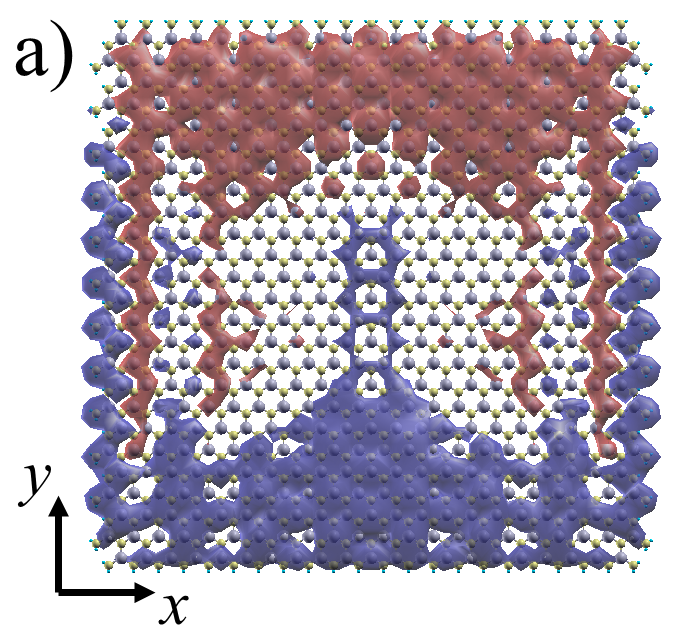}
\includegraphics[scale=0.23]{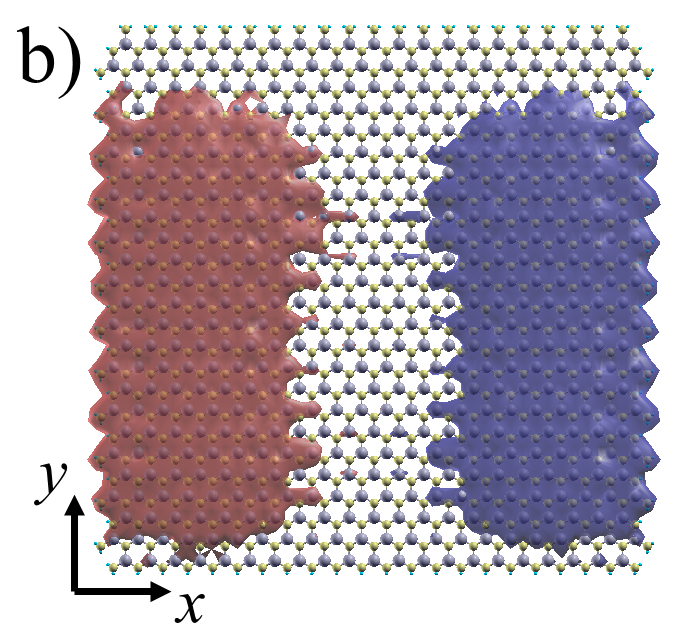}
\includegraphics[scale=0.23]{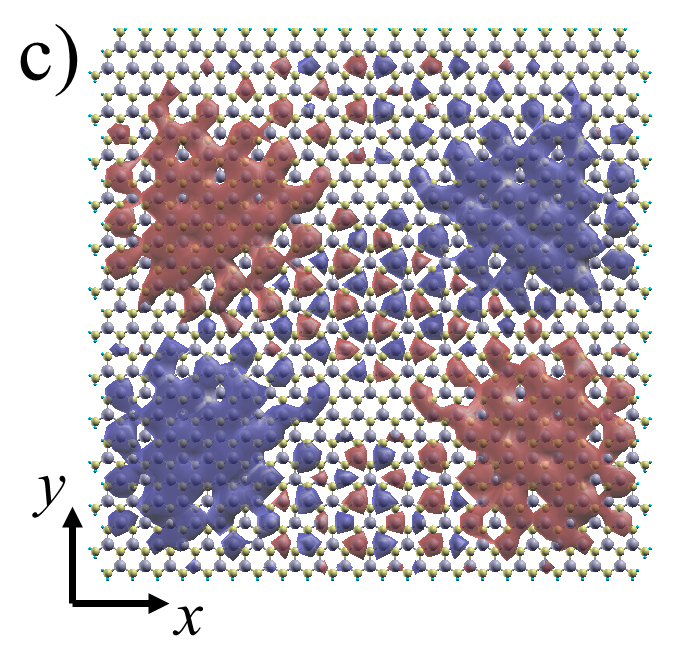}
\includegraphics[scale=0.215]{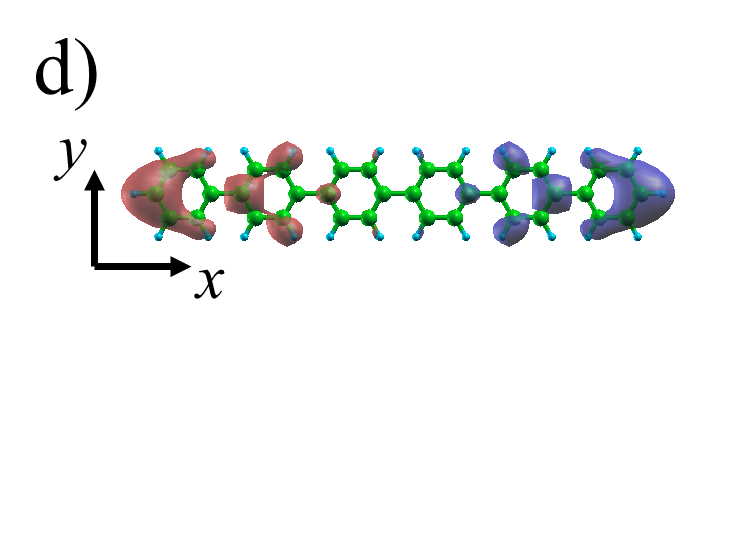}
      \caption{Top view on exciton transition densities $\rho$ for the MoS$_2$ nanoflake and the 6P molecule. Shown are isosurfaces embedded in the atomic lattice. The isosurfaces are obtained as the real part of $\rho_{\alpha 0}$ referring to the  (a) 1st, (b) 3rd  and  (c) 10th exciton level of the MoS$_2$ nanoflake as well as  (d) the single excitation  of the 6P molecule. Blue and red colored parts indicate positively and negatively charged areas, respectively.}
      \label{charge}
\end{figure}

In the gapped MoS$_2$ nanoflake, its excitonic energie and states can be given by Eq.\,\eqref{exciton}.
To characterize the excitonic properties,
we visualize the transition charge densities
 $\rho_{\alpha 0}$ of MoS$_2$ nanoflake by giving the value of real part of $q_v(\alpha 0)$ for excitonic level $\alpha$ at each atom $v$
(The imaginary part appears similar and is therefore not shown). 
Figure\,\ref{charge} shows transition densities for the 1st, 3rd and 10th excitonic levels.
Specifically, in Fig.\,\ref{charge}\,(a),  the transition densities  for the lowest exciton level have the energy of ${\cal E}_{\alpha=1}=1.709$ eV.
The related transition dipole moment is along the $y$-axis.
Similarly, the 2nd exciton level, ${\cal E}_{\alpha=2}=1.713$ eV, is characterized by a transition dipole moment along the $x$-axis as depicted in Fig.\,\ref{charge}\,(b).
We have ${\cal E}_{\alpha=10}=1.746$ eV, which has a quadrupole-like distribution in the $xy$-plane, as shown in Fig.\,\ref{charge}\,(c).
Obviously, we find that with the increase of the exciton energy, $\rho_{\alpha 0}$ displays more complex spatial patterns corresponding to the shapes of higher multipole moments.
Besides, all transition charge densities are symmetric distributions due to the bilateral symmetry of the nanoflake, and the associated transition dipole moments will be orientated in the $xy$-plane. 
For the organic component, the shape of 6P molecule transition charge densities (calculated by $q_u (eg)$) with the excitation energy $E_{eg}=4.014$ eV is illustrated in Fig.\,\ref{charge}\,(d),
as previously reported \cite{Plehn-2018-p27925}.

\subsection{Electron-hole pair approximation}

Since the excitation energy of the 6P molecule lies significantly above the bandgap of MoS$_2$ and even further above its lowest exciton energies, the resonant energy states in MoS$_2$ are not the strongly bound excitons but rather higher-energy, uncorrelated electron-hole pairs. 
%
Therefore,  we employ the electron-hole pair approximation, in which the Wannier-Mott excitons of the MoS$_2$ nanoflake are replaced by electron-hole pairs.
This treatment  is justified as it substantially reduces the computational cost while retaining physical relevance \cite{Ziemann-2014-p1203,Plehn-2018-p27925}.
Within this framework,  the Wannier-Mott exciton Hamiltonian $H_{\rm X}$ in Eq.\,\eqref{WMX} is simplified to the electron-hole pair Hamiltonian $H_{\rm e-h}$:
\begin{align}
H_{\rm X} \approx
H_{\rm e-h}=\sum_a E_a e_a^+ e_a - \sum_{\bar{a}} E_{\bar{a}}  h_{\bar{a}} ^+ h_{\bar{a}}\,.
\end{align}
Here, the exciton states in Eq.\,\eqref{x-states} are approximated as
$|\alpha\rangle \approx |\psi_{a \bar{a}}\rangle$,
and their corresponding energies are simply given by the single-particle energy difference of  electrons and holes, as ${\cal E}_{\alpha} \approx {\cal E}_{a\bar{a}}= E_a-E_{\bar{a}}$.
As a consequence, the EET coupling in Eq.\,\eqref{x-coupling} takes the following form 
\begin{align}
\label{aa-coupling}
V_{e0,\alpha g} \approx
V_{e0, a\bar{a}\,g} 
= \sum_{u,v} \frac{q_u (eg) q^*_v(a\bar{a})}{|\mathbf{R}_u - \mathbf{R}_v|}\,,
\end{align}
where $q^*_v(a\bar{a})$ indicates the transition charges associated with electron-hole pairs.
Although this approximation might lead to a lower EET coupling \cite{Ziemann-2014-p1203,Plehn-2018-p27925},
it still gives reasonable results when the considered electron-hole pair energies are far away from the bandgap.

\begin{figure}[!ht] 
	\centering  
	\includegraphics[scale=1.05]{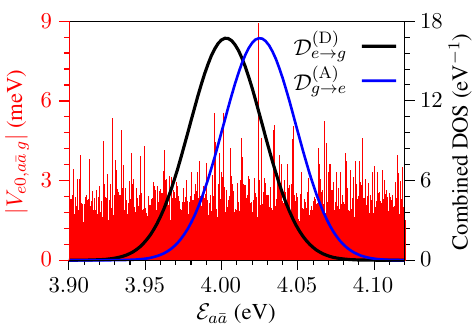} 
	\caption{Absolute values of the EET coupling $V_{e0, a\bar{a}\,g}$ between  MoS$_2$ electron-hole pairs and the 6P molecule  (red sticks) drawn versus energies of the electron-hole pairs. 
	The molecule is placed at the center of the MoS$_2$ nanoflake with a minimal distance of 2.0 {\AA}.
	 The black (blue) solid line is the combined DOS $\mathcal{D}^{(\text{D})}_{e\to g}$ ($\mathcal{D}^{(\text{A})}_{g\to e}$) 
	 from  Eq.\,\eqref{D2} (Eq.\,\eqref{D1}).} 
	\label{sto} 
\end{figure}

In addition,
after sorting the eigenenergies ($E_a$ and $E_{\bar{a}}$) of the  tight-binding Hamiltonian, 4081 occupied VB-like states and 2378 unoccupied CB-like states of the 70 {\AA}  MoS$_2$ nanoflake are obtained.
To further reduce the computational complexity, only electron-hole pairs with energies near the molecular excitation energy are  retained.
Figure\,\ref{sto} displays the absolute Coulomb coupling $|V_{e0, a\bar{a}\,g}|$ as a function of the electron-hole pair energy ${\cal E}_{a\bar{a}}$  (see red sticks), calculated with the surface-to-surface distance between surfaces of the MoS$_2$ nanoflake and the 6P molecule fixed at its minimum value of  2.0 {\AA}. 
The coupling strength remains below 9 meV throughout the investigated energy range.
This  relevant energy range is determined by the combined DOS also shown in Fig.\,\ref{sto}.
According to Eqs.\,\eqref{D1}) and \eqref{D2},  
the  de-excitation combined DOS of  the 6P molecule  $\mathcal{D}^{(D)}_{e\to g}$ 
  and excitation combined DOS $\mathcal{D}^{(A)}_{g\to e}$  peak at 3.992 eV and 4.036 eV, respectively.
It is required that electron-hole pair energies $\mathcal{E}_{a\bar{a}}$ overlap with molecular energies $E_{eg}\pm S_{eg}/2$, which contributes significantly to the energy transition.
Thus, we restrict our  calculations to the energy interval from 3.90 eV to 4.12 eV.

\subsection{Rates of EET}

\begin{figure}[ht]
      	\includegraphics[scale=1]{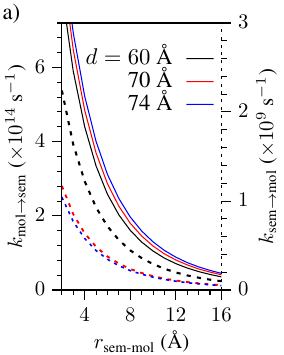} 
      	\includegraphics[scale=1]{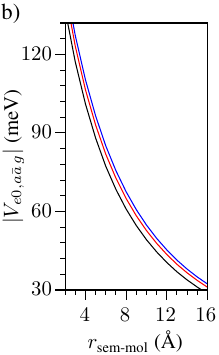} 
      \caption{EET rates $k_{\rm mol \to sem}$ (a) and EET couplings $|V_{e0, a\bar{a}\,g}|$ (b)  versus the distance along $z$-axis between the MoS$_2$ surface and the 6P molecule. The side lengths of MoS$_2$ nanoflakes are set to 60, 70 and 74 {\AA}.}
      \label{k-v}
\end{figure}

Next, we turn to the calculation of rates of EET.
Starting with the 6P molecule positioned at the center of MoS$_2$ layer, we vary the vertical separation along the $z$-axis.
Figure\,\ref{k-v}\,(a) presents the distance-dependent EET rates $k_{\rm mol \to sem}$ (Eq.\,\eqref{eq-rate}) and $k_{\rm sem \to mol}$ (Eq.\,\eqref{eq-rate2}) as a function of the surface-to-surface distance $r_{\rm sem-mol}$.  
We can see that both rates decay with the increasing separation.
Notably, $k_{\rm mol \to sem}$ exceeds $k_{\rm sem \to mol}$ by approximately five orders of magnitude, indicating that energy transfer from the molecule to the MoS$_2$ nanoflake  is the dominant process.
For $k_{\rm mol\to sem}$, as the distance increases from 2 {\AA} to 16 {\AA}, it drops from a maximum of $8.89\times10^{14}$ s$^{-1}$ to a minimum of $0.41\times10^{14}$ s$^{-1}$, corresponding to an increase in the excited-state lifetime of the molecule from about  1 fs to 14 fs.

The distance dependence of the EET rate is governed by the underlying Coulomb coupling.
Figure\,\ref{k-v}\,(b) shows the variation of the coupling strength $|V_{e0, a\bar{a}\,g}|$ with the distance $r_{\rm sem-mol}$.
At the minimal distance of 2 {\AA},
the coupling reaches a maximum value below 140 meV.
With the increase of the distance, a decrease of the coupling strength is obtained, directly dictating the trend observed for the EET rate, as expected from Eq.\,\eqref{aa-coupling}.

Furthermore, comparing systems with different scales of the nanoflake (lateral dimensions $d=60, 70$ and 74 {\AA}), a larger nanoflake size yields a stronger coupling  $|V_{e0, a\bar{a}\,g}|$, which consequently leads to a higher $k_{\rm mol\to sem}$.

\begin{figure}[ht] 
	\centering  
	\includegraphics[scale=0.35]{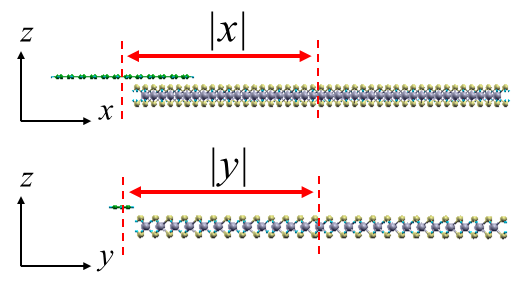} \\
   \includegraphics[scale=1]{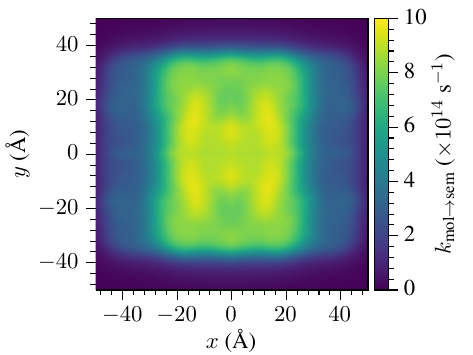} 
	\caption{EET rates versus the distance to the $xy$-plane for  70 {\AA} MoS$_2$ nanoflake. Distance to the $z$-axis is fixed at 2 {\AA}.}
	\label{k-x} 
\end{figure}

Finally, we examine how the EET rates vary with the lateral position of the 6P molecule in the $xy$-plane, as schematically illustrated  in the upper panel of Fig.\,\ref{k-x} for the 70 {\AA} MoS$_2$ square.
The vertical separation along the $z$-axis is also fixed at the minimal value of 2 {\AA}.
The calculating results are shown in the lower panel of Fig.\,\ref{k-x}.
We find that the spatial variation of the rate exhibits a bilaterally symmetric yet non-uniform, which originates from the underlying distribution of electron-hole states within the nanoflake.
Obviously, when the molecule is positioned within a central region approximately $\pm$35 {\AA} along both the $x$- and $y$-axis, the EET rate reaches a maximum on the order of 10$^{14}$ s$^{-1}$.
Beyond this central area, the rate decreases radically as the lateral displacement increases, falling to negligible values at the periphery of this nanoflake. 
%

%
%
\section{Conclusions}
\label{sec4}

In this work, we have theoretically investigated excitation energy transfer (EET) in an inorganic/organic nanohybrid system consisting of a single para-sexiphenyl (6P) molecule and a H-passivated MoS$_2$ nanoflake. 
The electronic structure of the finite-sized  MoS$_2$ nanoflake is described using an 11-band tight-binding model, with edge states removed via H passivation to recover a well-defined bandgap.
Wannier-Mott-like excitons in MoS$_2$ nanoflake are treated within a configuration-interaction scheme, and for computational efficiency, the exciton energies and exciton states are approximated by the uncorrelated electron-hole pair energies.
In addition, the first excited states of the 6P molecule have already been simulated in Ref.\,\citenum{Plehn-2018-p27925}, and the atomic-centered transition charges of the 6P molecule are used in this work.

The EET rate expressions,  evaluated within the Fermi golden-rule framework, are constituted by the MoS$_2$-6P transfer coupling, the thermal distribution, and the energetic resonance.
For the considered electron-hole pairs of the 70 {\AA} MoS$_2$ layer, the individual transfer couplings are all less than 10 meV, and decrease even further as the  surface-to-surface distances between the MoS$_2$ nanoflake and the 6P molecule increase. 
Due to the weak coupling strength,  the resulting EET rate from the molecule to the nanoflake drops sharply with separation: 
at the minimal distance of 2 {\AA}, 
the decay time of an excited 6P molecule is about 1 fs, whereas it increases substantially at larger separations.
A larger size of MoS$_2$ nanoflake causes  stronger Coulomb couplings and consequently higher EET rates.
We also examined the lateral-position dependence of the EET rate.
When the 6P molecule moves within the plane parallel to the MoS$_2$ surface, the rate remains the order of $10^{14}$ s$^{-1}$, beyond which it decreases rapidly.
The spatial variation reflects the symmetry of the nanoflake and the distribution of electron-hole states.
This strong distance and position dependence aligns with experimental trends observed in other TMDC–molecule hybrid systems \cite{Park-2019-p109,Gu-2018-p100,Gaur-2019-p173103}, which are also within the scope of application of our theory.
In summary, our analysis provides a quantitative theoretical description of non-contact energy transfer in a model organic-inorganic heterostructure, highlighting the roles of Coulomb coupling, spectral overlap, and interfacial geometry.
 The theoretical framework presented here holds promise for extension to EET study within few-layer MoS$_2$-based inorganic/organic heterostructures, where unique properties of few-layer MoS$_2$, such as indirect bandgap and interlayer exciton states, render it a compelling subject for future research endeavors.

%
%


\begin{acknowledgments}
We acknowledge the support from the National Natural Science Foundation of China through project no. 21961132023 (L.Wang).
\end{acknowledgments}

\bibliography{ref} 

\end{document}